\documentclass[aps,showpacs,twocolumn,amsmath,amssymb,superscriptaddress]{revtex4-1}
\usepackage[colorlinks=true, citecolor=blue, urlcolor=blue ]{hyperref}
\usepackage{graphicx}
\usepackage{subfigure}
\usepackage{grffile}
\usepackage{bm}
\usepackage[version=3]{mhchem}
\usepackage{color}
\usepackage{hyperref}
\usepackage[normalem]{ulem}

\newcommand{\abs}[1]{\ensuremath{\left| #1 \right|}}

\newcommand{\eval}[3]{\langle#1\vert#2\vert#3\rangle}

\newcommand{\1}{\mbox{\bf 1}}

\begin{document}

\title{The single-particle spectral function of the extended Peierls-Hubbard model at half-filling and quarter-filling}

\author{Ren-He Xu}
\affiliation{Department of Applied Physics $\&$ MIIT Key Laboratory of Semiconductor Microstructure and Quantum Sensing, Nanjing University of Science and Technology, Nanjing 210094, China}

\author{Hantao Lu}
\affiliation{School of Physical Science and Technology, Lanzhou University, Lanzhou 730000, China}
\affiliation{Lanzhou Center for Theoretical Physics $\&$ Key Laboratory of Theoretical
Physics of Gansu Province, Lanzhou University, Lanzhou 730000, China}

\author{Takami Tohyama}
\affiliation{Department of Applied Physics, Tokyo University of Science, Tokyo 125-8585, Japan}

\author{Can Shao}
\email{shaocan@njust.edu.cn}
\affiliation{Department of Applied Physics $\&$ MIIT Key Laboratory of Semiconductor Microstructure and Quantum Sensing, Nanjing University of Science and Technology, Nanjing 210094, China}

\date{\today}

\begin{abstract}
By utilizing the twisted boundary conditions in the exact diagonalization method, we investigate the single-particle spectral function of the extended Peierls-Hubbard model at both half-filling and quarter filling. In one-dimensional (1D) interacting systems, the spin-charge separation can typically be identified in the single-particle spectral function by observing the distinct spinon and holon bands. At half filling, starting from the pure 1D Hubbard model with the on-site interaction $U=10$, we observe that the band structure indicative of the spin-charge separation gradually transitions to four individual bands as the Peierls instability $\delta$ increases. At $U=10$ and $\delta=0.2$ where the spin-charge separation is still observable, increasing the nearest-neighbor interaction $V$ can drive the system to a charge-density-wave (CDW) state when $V\gtrsim U/2$, without the obeservation of spinon and holon bands.
At quarter-filling, on the other hand, the ground state of Peierls-Hubbard model manifests an antiferromagnetic Mott insulator in units of dimers. Increasing $U$ results in only a very small gap in the single-particle spectrum because even for $U=+\infty$, with the model transforming into a noninteracting half-filled dimerized tight-binding model, its gap determined by the Peierls instability $\delta$ remains small. Conversely, increasing $V$ can effectively open the single-particle gap and make the spinon and holon bands more prominent.
\end{abstract}


\maketitle

\section{Introduction}
\label{sec1}
Distinct from the Fermi liquid theory that describes the low-energy physics of many-particle systems based on the quasiparticle language, the low-energy charge and spin excitations of one-dimensional (1D) interacting systems belong to another paradigm referred to as the Tomonaga-Luttinger liquid (TLL)~\cite{Tomonaga50, Luttinger63, Haldane_1981, Voit_1993}. It is predicted by TLL that interactions between fermions in 1D systems can induce two separated collective excitations of electrons, known as `spinons' and `holons'. This phenomenon is well-known as spin-charge separation and has been widely investigated in experiments\cite{Kim96, Kim97, Fujisawa99, Segovia99, Auslaender02, Auslaender05, Jompol09, Senaratne22, Claessen02, Sing03, Kim2006}. Among these experiments, the angle-resolved photoelectron spectroscopy (ARPES) provides significant clues to its existence, with a direct distinction of the spinon and holon bands\cite{Kim96, Kim97, Fujisawa99, Claessen02, Sing03, Kim2006}. Theoretical simulations of ARPES in 1D interacting systems are typically based on the single-particle spectral function of 1D t-J and Hubbard models, which align well with the experimental results for spinon and holon bands~\cite{Sorella92, Penc95, Penc96, Penc97, Kim97, TOHYAMA98, Aichhorn04, Shao20}.

Another 1D interacting system, the Peierls-Hubbard model with both on-site interaction $U$ and bond dimerization $\delta$, has also attract significant attention due to the potential formation of a topological phase. In this paper, we focus on the single-particle spectral function of this model, incorporating the nearest-neighbor interaction $V$, and refer to it as the extended Peierls-Hubbard model. At half-filling, its ground-state phase diagram in the parameter space of ($U$, $V$) with explicit bond dimerization $\delta=0.2$ is studied, with the phase transition from Peierls insulator (PI) to charge-density-wave (CDW) state~\cite{Tsuchiizu04, Ejima16}. Although different tricritical points are given by Ref.~\cite{Tsuchiizu04} using a perturbative approach and Ref.~\cite{Ejima16} using the density-matrix renormalization group method, both proposing a continuous transition in the weak coupling regime and a first-order transition in the strong coupling regime. At quarter-filling, on the other hand, it is suggested that the ground state of the Peierls-Hubbard model is an antiferromagnetic Mott insulator in dimer units~\cite{Le2020} and is relevant for describing certain charge-transfer salts~\cite{Pedron94, Nishimoto2000, Shibata01, Tsuchiizu01, Penc94, Mila95, Favand96}. While for the extended Peierls-Hubbard model at quarter-filling, the main features of the optical conductivity spectrum are found to be determined by dimerization and nearest-neighbor repulsion $V$.~\cite{Benthien2005}.

In this paper, we present the results of the single-particle spectral function in the 1D extended Peierls-Hubbard model at both half-filling and quarter-filling, using the exact diagonalization (ED) method with twisted boundary conditions. At half-filling, we observe the evolution of the spectral function as the dimerization strength $\delta$ is increased from the pure Hubbard model with $U=10$. We find the formation of a hybridization gap and the suppression of spin-charge separation, characterized by the disappearance of holon and doublon bands. When keeping $U=10$ and $\delta=0.2$, increasing $V$ slightly decreases the single-particle gap until the system transforms into the CDW state, where the single-particle gap rapidly increase and the band structure of spin-charge separation is not observable.
At quarter-filling, we enlarge the interaction $U$ starting from a pure dimerized chain and observe a split of the energy band at Fermi level, indicating the formation of an antiferromagnetic Mott insulator in dimer units. The single-particle gap remains very small even for large interaction $U$; however, the interaction $V$ rapidly enlarges the gap, signifying the gap size is primarily determined by the nearest-neighbor interaction. Additionally, the band structure of spin-charge separation becomes more prominent as $V$ increases.

The rest of the paper is organized as follows: In Sec.~\ref{sec_model}, we introduce the model and method to obtain the single-particle spectral function. The analysis of our results both at half-filling and quarter-filling is shown in Sec.~\ref{sec_result}, and a conclusion is given in Sec.~\ref{sec_conclusion}.

\section{Model and Measurement}\label{sec_model}
We write the 1D extended Peierls-Hubbard model as
\begin{eqnarray}
H=H_k+H_I,
\label{H}
\end{eqnarray}
with the kinetic term
\begin{eqnarray}
H_k=-\sum_{i,\sigma}\left(t_h(1+\delta(-1)^{i}) c^{\dagger}_{i,\sigma} c_{i+1,\sigma}+\text{H.c.}\right)
\label{H1}
\end{eqnarray}
and the interaction term
\begin{eqnarray}
H_I=U\sum_{i}n_{i,\uparrow}n_{i,\downarrow}+V\sum_{i}(n_{i}-1)(n_{i+1}-1).
\label{H2}
\end{eqnarray}
Here $c^{\dagger}_{i,\sigma}$ ($c_{i,\sigma}$) is the creation (annihilation) operator of an electron at site $i$ with spin $\sigma$. The number operator $n_{i,\sigma}=c^{\dagger}_{i,\sigma}c_{i,\sigma}$ and $n_{i}=n_{i,\uparrow}+n_{i,\downarrow}$. The hopping constants within and between dimers are $t_h(1+\delta)$ and $t_h(1-\delta)$, respectively. $U$ ($V$) represents the on-site (nearest-neighbor) Coulomb interaction. The inclusion of the $-1$ in the $V$ term accounts for the positioning of the Fermi level precisely at the interface between electron-addition and electron-removal bands. We focus on the model at half-filling and quarter-filling, and study the single-particle spectral function at zero temperature. In the main text, the lattice size $L$ for the half-filling case is set to $12$, while for quarter-filling we set $L=16$. It is important to note that the number of unit cells (or dimers) is $L/2$.

Combining the standard periodic boundary condition (PBC) and twisted boundary conditions (TBCs) in ED method, the single-particle spectral function $I(k,\omega)$ can be solved with more momenta~\cite{Tsutsui96, Tohyama04, Shao20}. To incorporate TBCs, we introduce the following substitution into Eq.(\ref{H1}):
\begin{eqnarray}
t_h(1+\delta(-1)^{i})c^{\dagger}_{i,\sigma}c_{i+1,\sigma}+\text{H.c.}&&\nonumber \\
\longrightarrow e^{\mathrm{i}\frac{\kappa}{2}}(t_h(1+\delta(-1)^{i})c^{\dagger}_{i,\sigma}c_{i+1,\sigma}+\text{H.c.}&&,
\label{eq:twist}
\end{eqnarray}
and we use the fraction $\frac{\kappa}{2}$ to reflect the fact that distance between neighbouring sites is half the distance between neighbouring dimers, which is essentially the unit cells. The Hamiltonian is subsequently $\kappa$-dependent ($H^{\kappa}$), as well as its energy eigenvalues $E^{\kappa}_m$ and eigenstates $\Psi_m^{\kappa}$. Here $\Psi_0^{\kappa}$ and $E^{\kappa}_0$ are the ground state and the corresponding ground-state energy, respectively. Under these conditions, the single-particle spectral function can be expressed as follows:
\begin{eqnarray}
I(k,\omega)=I_{+}(k,\omega)+I_{-}(k,\omega),
\label{A}
\end{eqnarray}
with the electron-addition spectral function
\begin{eqnarray}
&I_{+}(k,\omega)=\nonumber \\
&\sum\limits_{m,\sigma}|\eval{\Psi_m^{\kappa}}{c_{k_0,\sigma}^{\dag}}{\Psi_0^{\kappa}}|^{2} \delta(\omega-(E_m^{\kappa}-E_0^{\kappa})-\mu_{\kappa})
\label{A+}
\end{eqnarray}
and the electron-removal spectral function
\begin{eqnarray}
&I_{-}(k,\omega)=\nonumber \\ &\sum\limits_{m,\sigma}\abs{\eval{\Psi_m^{\kappa}}{c_{k_0,\sigma}}{\Psi_0^{\kappa}}}^{2} \delta(\omega+(E_m^{\kappa}-E_0^{\kappa})-\mu_{\kappa}).
\label{A-}
\end{eqnarray}
Here $k=k_0+\kappa$, with $k_0$ the allowed momenta in standard PBC, i.e., $k_0=\frac{2\pi l}{L/2}$ ($l=0, 1, \ldots, \frac{L}{2}-1$). $c_{k_0,\sigma}$ and $c^{\dag}_{k_0,\sigma}$ are the Fourier transformation of $c_{i,\sigma}$ and $c^{\dag}_{i,\sigma}$, respectively, with the site $i\in2l+1$. Here, $\kappa$ can be fine-tuned to obtain $I(k,\omega)$ with high momentum resolution. The chemical potential $\mu_{\kappa}$ is set to be one half of the energy difference between the first ionization and affinity states of the system\cite{Tohyama04}. Eq.~(\ref{A+}) and Eq.~(\ref{A-}) are solved using the standard Lanczos technique with a spectral broadening factor $\eta=0.2$.


\section{Results of the single-particle spectral function}
\label{sec_result}

\subsection{Results at half-filling}\label{sec_half}

\begin{figure}
\centering
\includegraphics[width=0.5\textwidth]{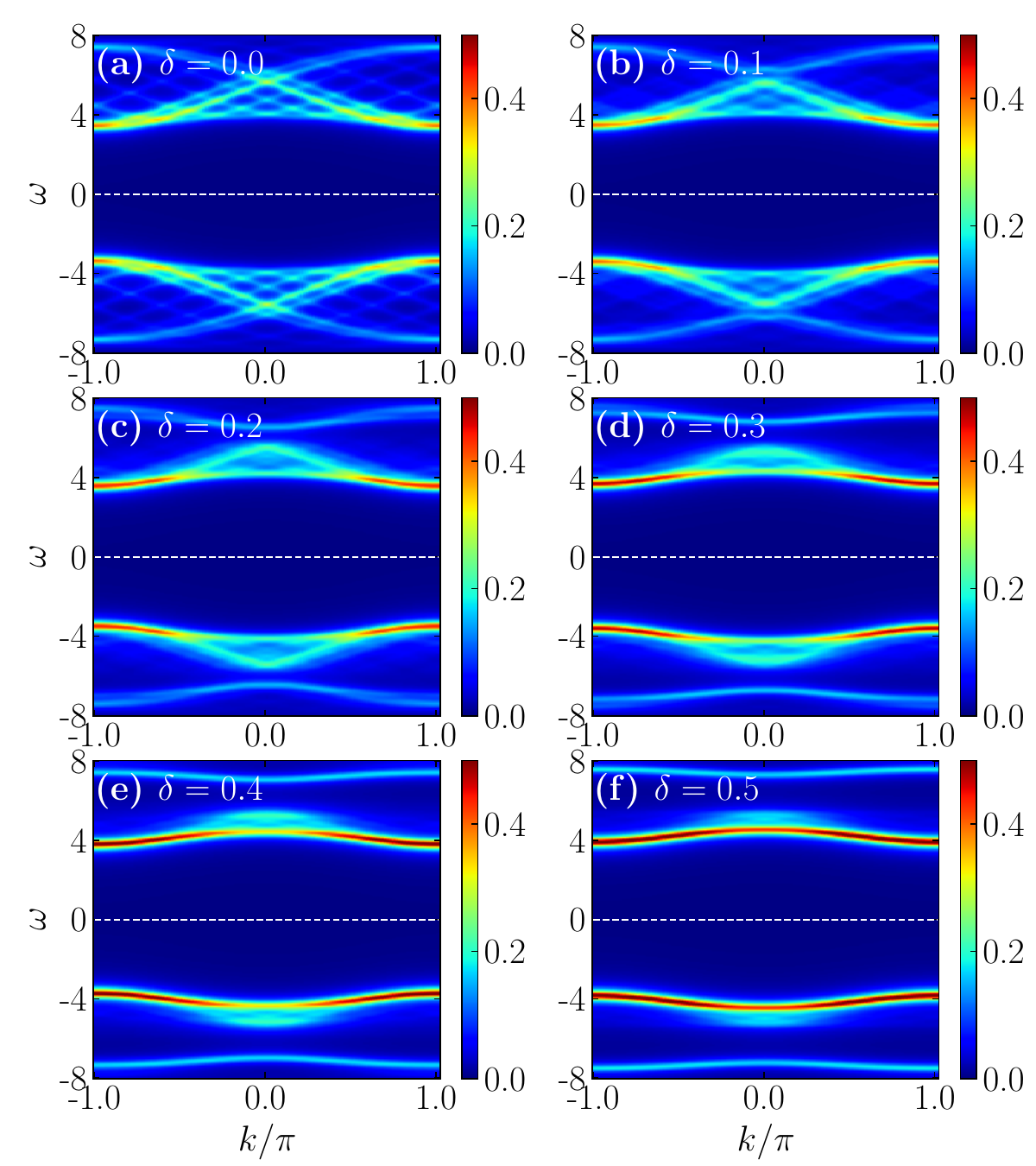}
\caption{(Color online) The single-particle spectral function $I(k,\omega)$ for the extended Peierls-Hubbard model at half-filling with $L=12$, $U=10$, $V=0$ and (a) $\delta=0.0$, (b) $\delta=0.1$, (c) $\delta=0.2$, (d) $\delta=0.3$, (e) $\delta=0.4$, and (f) $\delta=0.5$, respectively.}
\label{fig_1}
\end{figure}

For the extended Peierls-Hubbard model at half-filling with $U=10$ and $V=0$, we present the single-particle spectral function $I(k,\omega)$ for $\delta=0.0$, $\delta=0.1$, $\delta=0.2$, $\delta=0.3$, $\delta=0.4$ and $\delta=0.5$ in Figs.~\ref{fig_1}(a), \ref{fig_1}(b), \ref{fig_1}(c), \ref{fig_1}(d), \ref{fig_1}(e) and \ref{fig_1}(f), respectively.
The Fermi level (denoted by the dashed white line) is located in the middle between the lower and upper bands. In Fig.~\ref{fig_1}(a) with $\delta=0.0$, the system is actually the 1D Hubbard model without dimerization. However, to compare with the results for finite $\delta$, we calculate the single-particle spectral function in units of dimers (treat two neighboring sites as a unit cell) so that the first brillouin zone shrinks to half of its original size. For more details on their difference, refer to the single-particle spectral functions of the 1D Hubbard model in Appendix~\ref{appendix}. Spin-charge separation can be characterized from the single-particle spectral function, and a schematic view of the spinon and holon bands can be also found in Appendix~\ref{appendix}.

As $\delta$ increases in Fig.~\ref{fig_1}, the single-particle gap between the upper and lower bands increases slightly. On the other hand, two extra gaps emerge and widen at $k=0$ and $\omega\approx\pm6$, resulting in the holon band becoming ill-defined. These are the hybridization gap induced by the enhancement of bond dimerization. The interlaced stripes develop into four individual bands, and the spinon and holon branches gradually merge and become indistinguishable. This can be attributed to the fact that the combined effect of Coulomb repulsion and the Peierls instablity substantially localizes the mobility of electrons and thus suppress the charge excitation, i.e., holons.

\begin{figure}[t]
\centering
\includegraphics[width=0.5\textwidth]{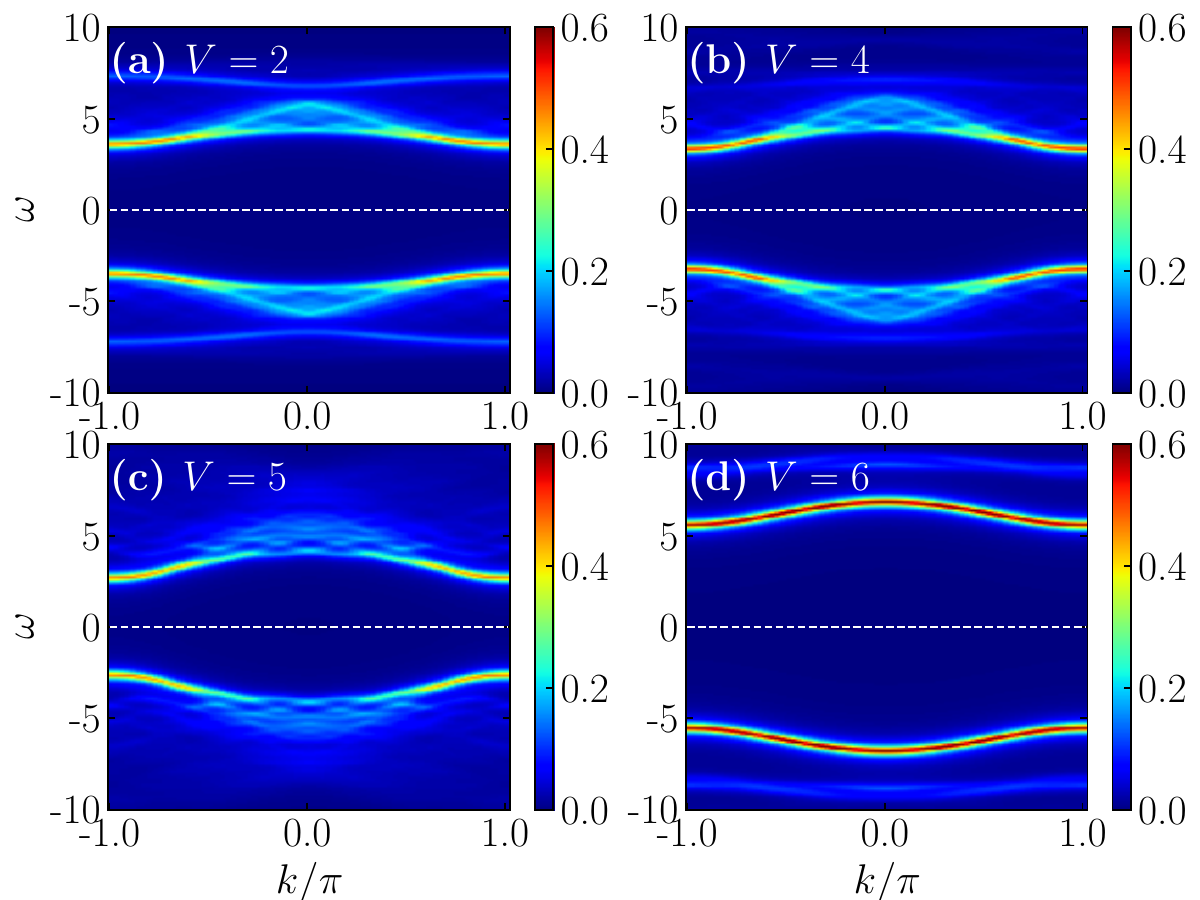}
\caption{(Color online) The single-particle spectral function $I(k,\omega)$ for the extended Peierls-Hubbard at half-filling with $L=12$, $\delta=0.2$, $U=10$ and (a) $V=1.0$, (b) $V=2.0$, (c) $V=3.0$, (d) $V=4.0$, (e) $V=5.0$, and (f) $V=6.0$, respectively.}
\label{fig_2}
\end{figure}

Setting $U=10$ and $\delta=0.2$ where the spinon and holon branches can still be observed in Fig.~\ref{fig_1}(c), we now introduce the nearest-neighbor interaction $V$ and the corresponding single-particle spectral functions are shown in Figs.~\ref{fig_2}(a), \ref{fig_2}(b), \ref{fig_2}(c), and \ref{fig_2}(d) with $V=2.0$, (b) $V=4.0$, (c) $V=5.0$, (d) $V=6.0$, respectively. Note that the range of $\omega$ in Fig.~\ref{fig_2} is $[-10,10]$, while in Fig.~\ref{fig_1} it is $[-8,8]$. According to the phase diagram based on the infinite density matrix renormalization group method in Ref.~\cite{Ejima16}, the nearest-neighbor interaction $V$ can drive the system from the Peierls insulator (PI) to the charge-density-wave (CDW) state when $V\gtrsim U/2$, under the condition of strong on-site interaction $U$. For the single-particle spectral function, increasing $V$ slightly decreases the single-particle gap in PI phase, as shown in Fig.~\ref{fig_2}(a), \ref{fig_2}(b) and \ref{fig_2}(c). While it rapidly increases the single-particle gap in the CDW phase (see Fig.~\ref{fig_2}(d)). These features are similar to the results of the single-particle spectrum in the 1D extended Hubbard model~\cite{Shao20}. In addition, the spinon and holon branches remain identifiable when $V=2$ and $V=4$ in the PI phase, but they become difficult to distinguish when $V=5$ which is around the critical point. When $V=6$ in the CDW phase, the band structure of spin-charge separation disappears and a four-band structure can be observed. This four-band structure also resembles the single-particle spectrum of the CDW state in the 1D extended Hubbard model~\cite{Shao20}. We speculate that the formation of CDW state not only localizes the holons, but also suppresses the spin excitations, i.e., spinons.

\subsection{Results at quarter-filling}\label{sec_quarter}

\begin{figure}
\centering
\includegraphics[width=0.5\textwidth]{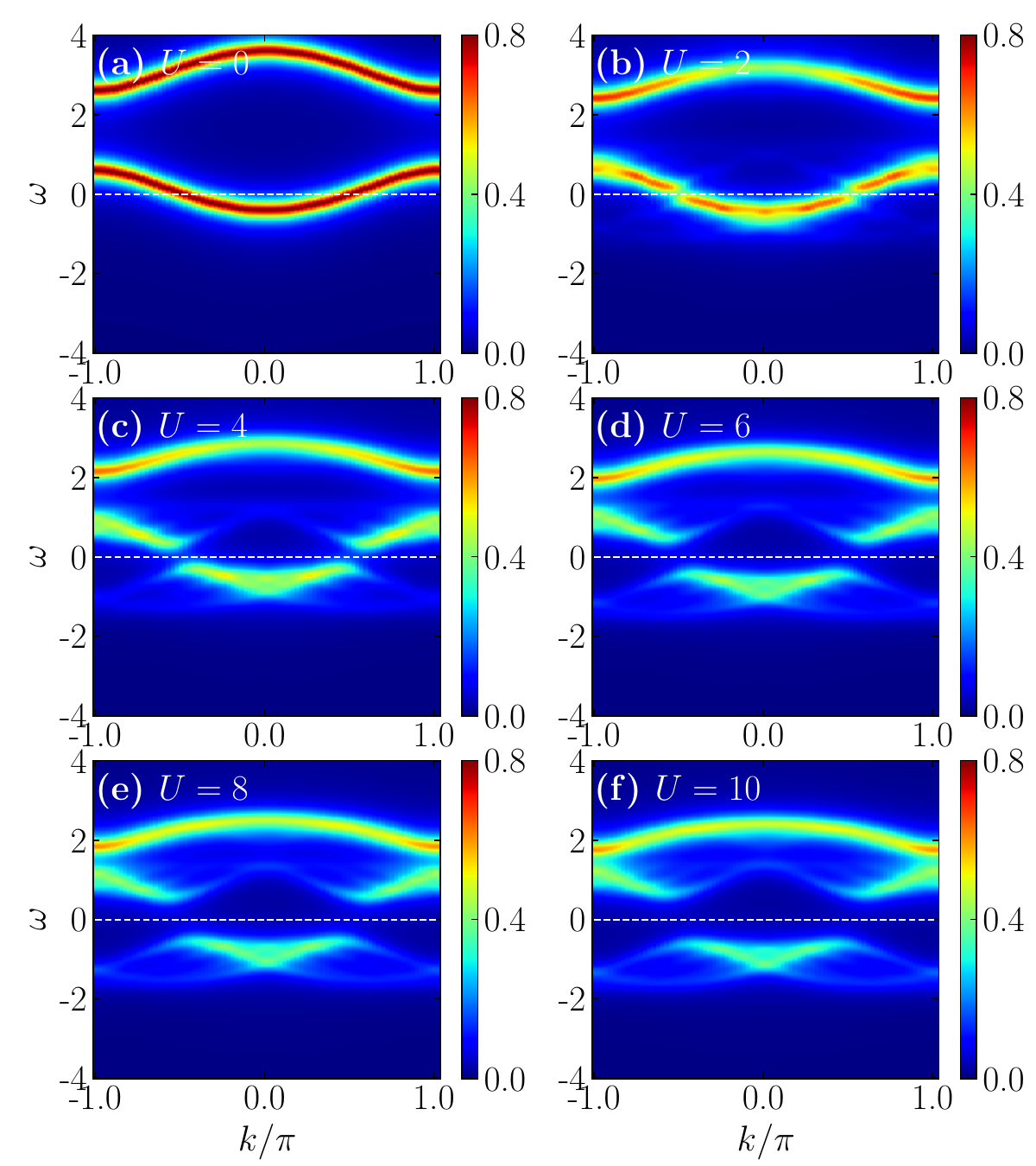}
\caption{(Color online) The single-particle spectral function $I(k,\omega)$ for the extended Peierls-Hubbard at quarter-filling with $L=16$, $\delta=0.5$, $V=0$ and (a) $U=0.0$, (b) $U=2.0$, (c) $U=4.0$, (d) $U=6.0$, (e) $U=8.0$, and (f) $U=10.0$, respectively.}
\label{fig_equilibrium}
\end{figure}

\begin{figure}[t]
\centering
\includegraphics[width=0.5\textwidth]{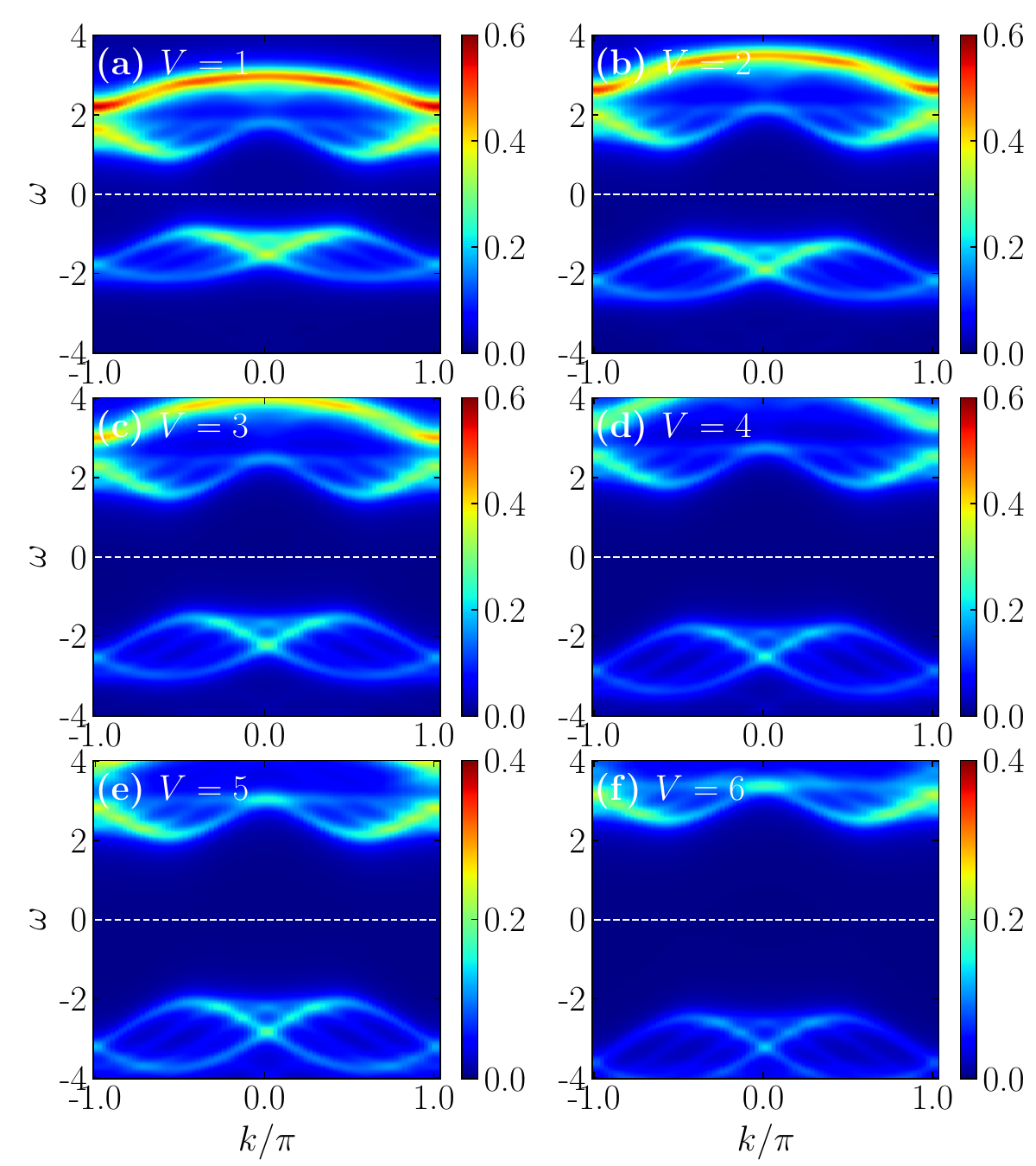}
\caption{(Color online) The single-particle spectral function $I(k,\omega)$ for the extended Peierls-Hubbard at quarter-filling with $L=16$, $\delta=0.5$, $U=10$, and (a) $V=1.0$, (b) $V=2.0$, (c) $V=3.0$, (d) $V=4.0$, (e) $V=5.0$, and (f) $V=6.0$, respectively.}
\label{fig_4}
\end{figure}

At half-filling, four individual bands (two below and two above the Fermi level) in the single-particle spectral function are induced by the combined effect of the interaction $U$ and the dimerization $\delta$. This observation motivates an examination of the system at quarter-filling.
We present the single-particle spectral function $I(k,\omega)$ for the quarter-filled Peierls-Hubbard model with $L=16$, $\delta=0.5$, $V=0$, and $U=0$, $2$, $4$, $6$, $8$, $10$ in Figs.~\ref{fig_equilibrium}(a), \ref{fig_equilibrium}(b), \ref{fig_equilibrium}(c), \ref{fig_equilibrium}(d), \ref{fig_equilibrium}(e), \ref{fig_equilibrium}(f), respectively.
In Fig.~\ref{fig_equilibrium}(a), the standard energy bands for the noninteracting tight-binding model in the dimerized chain can be observed, with the Fermi level (the white dashed line) positioned at the middle of the lower band due to the quarter-filled electrons. As $U$ increases, the lower band splits into two bands at the Fermi level, resulting in a three-band structure.  When $U\geq4$ in Figs.~\ref{fig_equilibrium}(c), \ref{fig_equilibrium}(d), \ref{fig_equilibrium}(e), \ref{fig_equilibrium}(f), the two newly generated bands around the Fermi level resemble the single-particle spectrum of the 1D half-filled Hubbard model on a $8$-site chain, see Fig.~\ref{fig_A1}(b) in Appendix~\ref{appendix}. The holon and spinon bands can be observed, although not very clearly. In addition, the two newly generated bands are invariant under the symmetric transformations $\omega\rightarrow-\omega$ and $k\rightarrow k+\pi$, which is consistent with the half-filled Hubbard model rather than the half-filled Peierls-Hubbard model.
This can be understood from the fact that the system of our quarter-filled Peierls-Hubbard model exhibits an antiferromagnetic Mott state in units of dimers, which is topologically trivial\cite{Le2020}, akin to the antiferromagnetic state of the Hubbard model in units of sites.

However, unlike the the single-particle spectral function of the Hubbard chain, the gap size of quarter-filled Peierls-Hubbard model remains very small compared to the values of $U$. This is because even for $U\rightarrow\infty$, the quarter-filled Peierls-Hubbard model transforms into the noninteracting half-filled tight-binding model on a dimerized chain. The single-particle spectral function of the latter model is identical to Fig.~\ref{fig_equilibrium}(a), but with the Fermi level positioned at the middle of the two bands. We find that the single-particle gap in Fig.~\ref{fig_equilibrium}(a), determined by the parameter $\delta$, remains small.

Setting $U=10$, we then present the single-particle spectral function $I(k,\omega)$ of the quarter-filled extended Peierls-Hubbard model with $V=1$, $2$, $3$, $4$, $5$ and $6$ in Figs.~\ref{fig_4}(a), \ref{fig_4}(b), \ref{fig_4}(c), \ref{fig_4}(d), \ref{fig_4}(e) and \ref{fig_4}(f), respectively. Note that the range of the color bars in Figs.~\ref{fig_4}(a), \ref{fig_4}(b), \ref{fig_4}(c) and \ref{fig_4}(d) is [0.0, 0.6], while that in Figs.~\ref{fig_4}(e) and \ref{fig_4}(f) is [0.0, 0.4]. This setting makes the spectra in Figs.~\ref{fig_4}(e) and \ref{fig_4}(f) more prominent. Notably, the nearest-neighbour interaction $V$ effectively increase the single-particle gap. This feature confirms the DMRG results in Ref.~\cite{Benthien2005} that the main features of their optical spectrum are determined by the dimerization and the nearest-neighbor repulsion. Additionally, as $V$ increases, the bands around the Fermi level more closely resemble the the single-particle spectrum of the 1D half-filled Hubbard model on the $8$-site chain (see Appendix~\ref{appendix}), with four ``stripes" easily distinguishable, especially for the band below the Fermi level. We speculate that interaction $V$ further enhance the dimerization of electrons and is beneficial to the robustness of the antiferromagnetic Mott state in units of dimers.

\section{Conclusion}\label{sec_conclusion}

To summarize, employing the twisted boundary conditions in the exact diagonalization method, we presented the outcomes of the single-particle spectral function with high momentum resolution in the 1D extended Peierls-Hubbard model at half-filling and quarter-filling. Starting with a half-filled Hubbard chain, we observe that increasing the dimerization strength $\delta$ leads to the emergence of a hybridization gap and disappearance of the spinon and holon bands. Increasing $V$ slghtly decreases the single-particle gap in the Peierls insulator phase and rapidly increase it in the charge-density-wave phase. At quarter-filling, ground state of 1D extended Peierls-Hubbard model with $V=0$ is an antiferromagnetic Mott insulator in dimer units. The single-particle gap is small even for large $U$, while the interaction $V$ rapidly widens the gap and makes the bands more similar to that of the 1D half-filled Hubbard model with half the lattice size.


\begin{acknowledgments}
C. S. acknowledges support from the National Natural Science Foundation of China (NSFC; Grant No. 12104229) and the Fundamental Research Funds for the Central Universities (Grant No. 30922010803).
T. T. is partly supported by the Japan Society for the Promotion of Science, KAKENHI (Grant No. 24K00560) from the Ministry of Education, Culture, Sports, Science, and Technology, Japan.
H. L. acknowledges support from the National Natural Science Foundation of China (NSFC; Grants No. 11474136, No. 11674139, and No. 11834005) and the Fundamental Research Funds for the Central Universities.
\end{acknowledgments}

\vskip 0.1in

\appendix

\section{$I(k,\omega)$ of the half-filled 1D Hubbard model} \label{appendix}

\begin{figure}[t]
\centering
\includegraphics[width=0.5\textwidth]{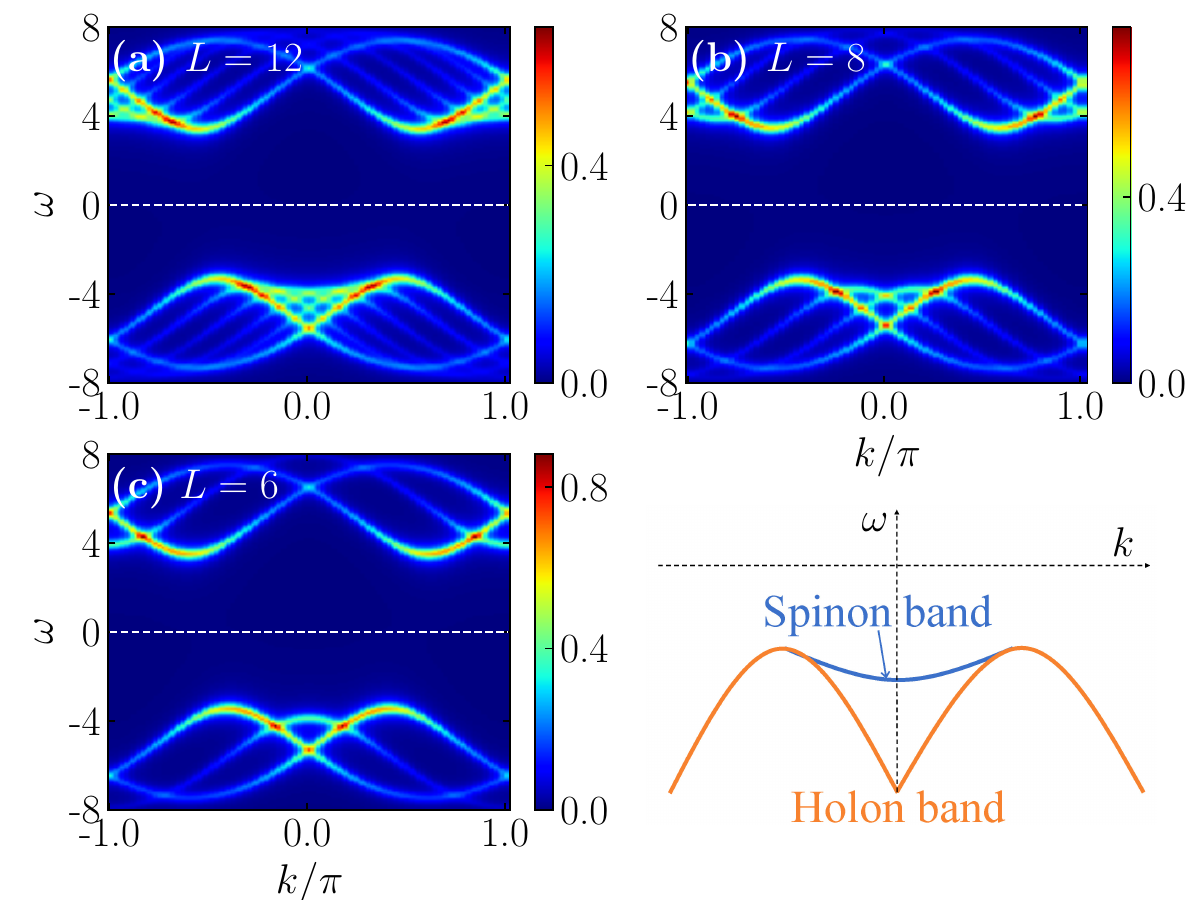}
\caption{(Color online) The single-particle spectral function $I(k,\omega)$ for the $U=10$ Hubbard model at half-filling with (a) $L=12$, (b) $L=8$ and (c) $L=6$. (d) Schematic view of the spinon and holon bands in thermodynamic limit.}
\label{fig_A1}
\end{figure}

In Fig.~\ref{fig_A1}(a), we present the single-particle spectral function for the 1D Hubbard model with lattice size $L=12$ in the original brillouin zone. We find that if the spectrum within $k\in[-\pi, -\pi/2]$ and $k\in[\pi/2, \pi]$ is folded into $k\in[-\pi/2,\pi/2]$, it becomes identical to the spectrum in Fig.~\ref{fig_1}(a), but with the rang of $k$ extended to $[-\pi,\pi]$. As a result, the lower and upper bands in Fig.~\ref{fig_1}(a) are invariant under the symmetric transformation of $\omega\leftrightarrow-\omega$ for the Peierls-Hubbard model. However, for the Hubbard model in Fig.~\ref{fig_A1}(a), they are invariant when the two symmetric transformations $\omega\leftrightarrow-\omega$ and $k\leftrightarrow k+\pi$ are satisfied. From both Fig.~\ref{fig_1}(a) and Fig.~\ref{fig_A1}(a), we can observe some striped bands due to the finite-size effect. We then present the single-particle spectral function of Hubbard model with lattice size $L=8$ and $L=6$ in Fig.~\ref{fig_A1}(b) and Fig.~\ref{fig_A1}(c). Compared to Fig.~\ref{fig_1}(a) with $6$ interlaced stripes in a band, number of the interlaced stripes in Fig.~\ref{fig_A1}(b) and Fig.~\ref{fig_A1}(c) are reduced to $4$ and $3$, respectively. The results with $L=10$ and $L=14$ can be found in Ref.~\cite{Shao20, Su_2023} and these stripes will develop into the spinon and holon branches as a result of the spin-charge separation in 1D interacting systems. A schematic view of the spinon and holon bands in thermodynamic limit are shown in Fig.~\ref{fig_A1}(d).


\begin{thebibliography}{36}%
\makeatletter
\providecommand \@ifxundefined [1]{%
 \@ifx{#1\undefined}
}%
\providecommand \@ifnum [1]{%
 \ifnum #1\expandafter \@firstoftwo
 \else \expandafter \@secondoftwo
 \fi
}%
\providecommand \@ifx [1]{%
 \ifx #1\expandafter \@firstoftwo
 \else \expandafter \@secondoftwo
 \fi
}%
\providecommand \natexlab [1]{#1}%
\providecommand \enquote  [1]{``#1''}%
\providecommand \bibnamefont  [1]{#1}%
\providecommand \bibfnamefont [1]{#1}%
\providecommand \citenamefont [1]{#1}%
\providecommand \href@noop [0]{\@secondoftwo}%
\providecommand \href [0]{\begingroup \@sanitize@url \@href}%
\providecommand \@href[1]{\@@startlink{#1}\@@href}%
\providecommand \@@href[1]{\endgroup#1\@@endlink}%
\providecommand \@sanitize@url [0]{\catcode `\\12\catcode `\$12\catcode
  `\&12\catcode `\#12\catcode `\^12\catcode `\_12\catcode `\%12\relax}%
\providecommand \@@startlink[1]{}%
\providecommand \@@endlink[0]{}%
\providecommand \url  [0]{\begingroup\@sanitize@url \@url }%
\providecommand \@url [1]{\endgroup\@href {#1}{\urlprefix }}%
\providecommand \urlprefix  [0]{URL }%
\providecommand \Eprint [0]{\href }%
\providecommand \doibase [0]{http://dx.doi.org/}%
\providecommand \selectlanguage [0]{\@gobble}%
\providecommand \bibinfo  [0]{\@secondoftwo}%
\providecommand \bibfield  [0]{\@secondoftwo}%
\providecommand \translation [1]{[#1]}%
\providecommand \BibitemOpen [0]{}%
\providecommand \bibitemStop [0]{}%
\providecommand \bibitemNoStop [0]{.\EOS\space}%
\providecommand \EOS [0]{\spacefactor3000\relax}%
\providecommand \BibitemShut  [1]{\csname bibitem#1\endcsname}%
\let\auto@bib@innerbib\@empty
\bibitem [{\citenamefont {Tomonaga}(1950)}]{Tomonaga50}%
  \BibitemOpen
  \bibfield  {author} {\bibinfo {author} {\bibfnamefont {S.-I.}\ \bibnamefont
  {Tomonaga}},\ }\href {\doibase 10.1143/ptp/5.4.544} {\bibfield  {journal}
  {\bibinfo  {journal} {Prog. Theor. Phys.}\ }\textbf {\bibinfo {volume} {5}},\
  \bibinfo {pages} {544} (\bibinfo {year} {1950})}\BibitemShut {NoStop}%
\bibitem [{\citenamefont {Luttinger}(1963)}]{Luttinger63}%
  \BibitemOpen
  \bibfield  {author} {\bibinfo {author} {\bibfnamefont {J.~M.}\ \bibnamefont
  {Luttinger}},\ }\href {\doibase 10.1063/1.1704046} {\bibfield  {journal}
  {\bibinfo  {journal} {J. Math. Phys.}\ }\textbf {\bibinfo {volume} {4}},\
  \bibinfo {pages} {1154} (\bibinfo {year} {1963})}\BibitemShut {NoStop}%
\bibitem [{\citenamefont {Haldane}(1981)}]{Haldane_1981}%
  \BibitemOpen
  \bibfield  {author} {\bibinfo {author} {\bibfnamefont {F.~D.~M.}\
  \bibnamefont {Haldane}},\ }\href {\doibase 10.1088/0022-3719/14/19/010}
  {\bibfield  {journal} {\bibinfo  {journal} {J. Phys. C: Solid State Phys.}\
  }\textbf {\bibinfo {volume} {14}},\ \bibinfo {pages} {2585} (\bibinfo {year}
  {1981})}\BibitemShut {NoStop}%
\bibitem [{\citenamefont {Voit}(1993)}]{Voit_1993}%
  \BibitemOpen
  \bibfield  {author} {\bibinfo {author} {\bibfnamefont {J.}~\bibnamefont
  {Voit}},\ }\href {\doibase 10.1088/0953-8984/5/44/020} {\bibfield  {journal}
  {\bibinfo  {journal} {J. Phys.: Condens. Matter}\ }\textbf {\bibinfo {volume}
  {5}},\ \bibinfo {pages} {8305} (\bibinfo {year} {1993})}\BibitemShut
  {NoStop}%
\bibitem [{\citenamefont {Kim}\ \emph {et~al.}(1996)\citenamefont {Kim},
  \citenamefont {Matsuura}, \citenamefont {Shen}, \citenamefont {Motoyama},
  \citenamefont {Eisaki}, \citenamefont {Uchida}, \citenamefont {Tohyama},\
  and\ \citenamefont {Maekawa}}]{Kim96}%
  \BibitemOpen
  \bibfield  {author} {\bibinfo {author} {\bibfnamefont {C.}~\bibnamefont
  {Kim}}, \bibinfo {author} {\bibfnamefont {A.~Y.}\ \bibnamefont {Matsuura}},
  \bibinfo {author} {\bibfnamefont {Z.-X.}\ \bibnamefont {Shen}}, \bibinfo
  {author} {\bibfnamefont {N.}~\bibnamefont {Motoyama}}, \bibinfo {author}
  {\bibfnamefont {H.}~\bibnamefont {Eisaki}}, \bibinfo {author} {\bibfnamefont
  {S.}~\bibnamefont {Uchida}}, \bibinfo {author} {\bibfnamefont
  {T.}~\bibnamefont {Tohyama}}, \ and\ \bibinfo {author} {\bibfnamefont
  {S.}~\bibnamefont {Maekawa}},\ }\href {\doibase 10.1103/PhysRevLett.77.4054}
  {\bibfield  {journal} {\bibinfo  {journal} {Phys. Rev. Lett.}\ }\textbf
  {\bibinfo {volume} {77}},\ \bibinfo {pages} {4054} (\bibinfo {year}
  {1996})}\BibitemShut {NoStop}%
\bibitem [{\citenamefont {Kim}\ \emph {et~al.}(1997)\citenamefont {Kim},
  \citenamefont {Shen}, \citenamefont {Motoyama}, \citenamefont {Eisaki},
  \citenamefont {Uchida}, \citenamefont {Tohyama},\ and\ \citenamefont
  {Maekawa}}]{Kim97}%
  \BibitemOpen
  \bibfield  {author} {\bibinfo {author} {\bibfnamefont {C.}~\bibnamefont
  {Kim}}, \bibinfo {author} {\bibfnamefont {Z.-X.}\ \bibnamefont {Shen}},
  \bibinfo {author} {\bibfnamefont {N.}~\bibnamefont {Motoyama}}, \bibinfo
  {author} {\bibfnamefont {H.}~\bibnamefont {Eisaki}}, \bibinfo {author}
  {\bibfnamefont {S.}~\bibnamefont {Uchida}}, \bibinfo {author} {\bibfnamefont
  {T.}~\bibnamefont {Tohyama}}, \ and\ \bibinfo {author} {\bibfnamefont
  {S.}~\bibnamefont {Maekawa}},\ }\href {\doibase 10.1103/PhysRevB.56.15589}
  {\bibfield  {journal} {\bibinfo  {journal} {Phys. Rev. B}\ }\textbf {\bibinfo
  {volume} {56}},\ \bibinfo {pages} {15589} (\bibinfo {year}
  {1997})}\BibitemShut {NoStop}%
\bibitem [{\citenamefont {Fujisawa}\ \emph {et~al.}(1999)\citenamefont
  {Fujisawa}, \citenamefont {Yokoya}, \citenamefont {Takahashi}, \citenamefont
  {Miyasaka}, \citenamefont {Kibune},\ and\ \citenamefont
  {Takagi}}]{Fujisawa99}%
  \BibitemOpen
  \bibfield  {author} {\bibinfo {author} {\bibfnamefont {H.}~\bibnamefont
  {Fujisawa}}, \bibinfo {author} {\bibfnamefont {T.}~\bibnamefont {Yokoya}},
  \bibinfo {author} {\bibfnamefont {T.}~\bibnamefont {Takahashi}}, \bibinfo
  {author} {\bibfnamefont {S.}~\bibnamefont {Miyasaka}}, \bibinfo {author}
  {\bibfnamefont {M.}~\bibnamefont {Kibune}}, \ and\ \bibinfo {author}
  {\bibfnamefont {H.}~\bibnamefont {Takagi}},\ }\href {\doibase
  10.1103/PhysRevB.59.7358} {\bibfield  {journal} {\bibinfo  {journal} {Phys.
  Rev. B}\ }\textbf {\bibinfo {volume} {59}},\ \bibinfo {pages} {7358}
  (\bibinfo {year} {1999})}\BibitemShut {NoStop}%
\bibitem [{\citenamefont {Segovia}\ \emph {et~al.}(1999)\citenamefont
  {Segovia}, \citenamefont {Purdie}, \citenamefont {Hengsberger},\ and\
  \citenamefont {Baer}}]{Segovia99}%
  \BibitemOpen
  \bibfield  {author} {\bibinfo {author} {\bibfnamefont {P.}~\bibnamefont
  {Segovia}}, \bibinfo {author} {\bibfnamefont {D.}~\bibnamefont {Purdie}},
  \bibinfo {author} {\bibfnamefont {M.}~\bibnamefont {Hengsberger}}, \ and\
  \bibinfo {author} {\bibfnamefont {Y.}~\bibnamefont {Baer}},\ }\href {\doibase
  10.1038/990052} {\bibfield  {journal} {\bibinfo  {journal} {Nature}\ }\textbf
  {\bibinfo {volume} {402}},\ \bibinfo {pages} {504} (\bibinfo {year}
  {1999})}\BibitemShut {NoStop}%
\bibitem [{\citenamefont {Auslaender}\ \emph {et~al.}(2002)\citenamefont
  {Auslaender}, \citenamefont {Yacoby}, \citenamefont {de~Picciotto},
  \citenamefont {Baldwin}, \citenamefont {Pfeiffer},\ and\ \citenamefont
  {West}}]{Auslaender02}%
  \BibitemOpen
  \bibfield  {author} {\bibinfo {author} {\bibfnamefont {O.~M.}\ \bibnamefont
  {Auslaender}}, \bibinfo {author} {\bibfnamefont {A.}~\bibnamefont {Yacoby}},
  \bibinfo {author} {\bibfnamefont {R.}~\bibnamefont {de~Picciotto}}, \bibinfo
  {author} {\bibfnamefont {K.~W.}\ \bibnamefont {Baldwin}}, \bibinfo {author}
  {\bibfnamefont {L.~N.}\ \bibnamefont {Pfeiffer}}, \ and\ \bibinfo {author}
  {\bibfnamefont {K.~W.}\ \bibnamefont {West}},\ }\href {\doibase
  10.1126/science.1066266} {\bibfield  {journal} {\bibinfo  {journal}
  {Science}\ }\textbf {\bibinfo {volume} {295}},\ \bibinfo {pages} {825}
  (\bibinfo {year} {2002})}\BibitemShut {NoStop}%
\bibitem [{\citenamefont {Auslaender}\ \emph {et~al.}(2005)\citenamefont
  {Auslaender}, \citenamefont {Steinberg}, \citenamefont {Yacoby},
  \citenamefont {Tserkovnyak}, \citenamefont {Halperin}, \citenamefont
  {Baldwin}, \citenamefont {Pfeiffer},\ and\ \citenamefont
  {West}}]{Auslaender05}%
  \BibitemOpen
  \bibfield  {author} {\bibinfo {author} {\bibfnamefont {O.~M.}\ \bibnamefont
  {Auslaender}}, \bibinfo {author} {\bibfnamefont {H.}~\bibnamefont
  {Steinberg}}, \bibinfo {author} {\bibfnamefont {A.}~\bibnamefont {Yacoby}},
  \bibinfo {author} {\bibfnamefont {Y.}~\bibnamefont {Tserkovnyak}}, \bibinfo
  {author} {\bibfnamefont {B.~I.}\ \bibnamefont {Halperin}}, \bibinfo {author}
  {\bibfnamefont {K.~W.}\ \bibnamefont {Baldwin}}, \bibinfo {author}
  {\bibfnamefont {L.~N.}\ \bibnamefont {Pfeiffer}}, \ and\ \bibinfo {author}
  {\bibfnamefont {K.~W.}\ \bibnamefont {West}},\ }\href {\doibase
  10.1126/science.1107821} {\bibfield  {journal} {\bibinfo  {journal}
  {Science}\ }\textbf {\bibinfo {volume} {308}},\ \bibinfo {pages} {88}
  (\bibinfo {year} {2005})}\BibitemShut {NoStop}%
\bibitem [{\citenamefont {Jompol}\ \emph {et~al.}(2009)\citenamefont {Jompol},
  \citenamefont {Ford}, \citenamefont {Griffiths}, \citenamefont {Farrer},
  \citenamefont {Jones}, \citenamefont {Anderson}, \citenamefont {Ritchie},
  \citenamefont {Silk},\ and\ \citenamefont {Schofield}}]{Jompol09}%
  \BibitemOpen
  \bibfield  {author} {\bibinfo {author} {\bibfnamefont {Y.}~\bibnamefont
  {Jompol}}, \bibinfo {author} {\bibfnamefont {C.~J.~B.}\ \bibnamefont {Ford}},
  \bibinfo {author} {\bibfnamefont {J.~P.}\ \bibnamefont {Griffiths}}, \bibinfo
  {author} {\bibfnamefont {I.}~\bibnamefont {Farrer}}, \bibinfo {author}
  {\bibfnamefont {G.~A.~C.}\ \bibnamefont {Jones}}, \bibinfo {author}
  {\bibfnamefont {D.}~\bibnamefont {Anderson}}, \bibinfo {author}
  {\bibfnamefont {D.~A.}\ \bibnamefont {Ritchie}}, \bibinfo {author}
  {\bibfnamefont {T.~W.}\ \bibnamefont {Silk}}, \ and\ \bibinfo {author}
  {\bibfnamefont {A.~J.}\ \bibnamefont {Schofield}},\ }\href {\doibase
  10.1126/science.1171769} {\bibfield  {journal} {\bibinfo  {journal}
  {Science}\ }\textbf {\bibinfo {volume} {325}},\ \bibinfo {pages} {597}
  (\bibinfo {year} {2009})}\BibitemShut {NoStop}%
\bibitem [{\citenamefont {Senaratne}\ \emph {et~al.}(2022)\citenamefont
  {Senaratne}, \citenamefont {Cavazos-Cavazos}, \citenamefont {Wang},
  \citenamefont {He}, \citenamefont {Chang}, \citenamefont {Kafle},
  \citenamefont {Pu}, \citenamefont {Guan},\ and\ \citenamefont
  {Hulet}}]{Senaratne22}%
  \BibitemOpen
  \bibfield  {author} {\bibinfo {author} {\bibfnamefont {R.}~\bibnamefont
  {Senaratne}}, \bibinfo {author} {\bibfnamefont {D.}~\bibnamefont
  {Cavazos-Cavazos}}, \bibinfo {author} {\bibfnamefont {S.}~\bibnamefont
  {Wang}}, \bibinfo {author} {\bibfnamefont {F.}~\bibnamefont {He}}, \bibinfo
  {author} {\bibfnamefont {Y.-T.}\ \bibnamefont {Chang}}, \bibinfo {author}
  {\bibfnamefont {A.}~\bibnamefont {Kafle}}, \bibinfo {author} {\bibfnamefont
  {H.}~\bibnamefont {Pu}}, \bibinfo {author} {\bibfnamefont {X.-W.}\
  \bibnamefont {Guan}}, \ and\ \bibinfo {author} {\bibfnamefont {R.~G.}\
  \bibnamefont {Hulet}},\ }\href {\doibase 10.1126/science.abn1719} {\bibfield
  {journal} {\bibinfo  {journal} {Science}\ }\textbf {\bibinfo {volume}
  {376}},\ \bibinfo {pages} {1305} (\bibinfo {year} {2022})}\BibitemShut
  {NoStop}%
\bibitem [{\citenamefont {Claessen}\ \emph {et~al.}(2002)\citenamefont
  {Claessen}, \citenamefont {Sing}, \citenamefont {Schwingenschl\"ogl},
  \citenamefont {Blaha}, \citenamefont {Dressel},\ and\ \citenamefont
  {Jacobsen}}]{Claessen02}%
  \BibitemOpen
  \bibfield  {author} {\bibinfo {author} {\bibfnamefont {R.}~\bibnamefont
  {Claessen}}, \bibinfo {author} {\bibfnamefont {M.}~\bibnamefont {Sing}},
  \bibinfo {author} {\bibfnamefont {U.}~\bibnamefont {Schwingenschl\"ogl}},
  \bibinfo {author} {\bibfnamefont {P.}~\bibnamefont {Blaha}}, \bibinfo
  {author} {\bibfnamefont {M.}~\bibnamefont {Dressel}}, \ and\ \bibinfo
  {author} {\bibfnamefont {C.~S.}\ \bibnamefont {Jacobsen}},\ }\href {\doibase
  10.1103/PhysRevLett.88.096402} {\bibfield  {journal} {\bibinfo  {journal}
  {Phys. Rev. Lett.}\ }\textbf {\bibinfo {volume} {88}},\ \bibinfo {pages}
  {096402} (\bibinfo {year} {2002})}\BibitemShut {NoStop}%
\bibitem [{\citenamefont {Sing}\ \emph {et~al.}(2003)\citenamefont {Sing},
  \citenamefont {Schwingenschl\"ogl}, \citenamefont {Claessen}, \citenamefont
  {Blaha}, \citenamefont {Carmelo}, \citenamefont {Martelo}, \citenamefont
  {Sacramento}, \citenamefont {Dressel},\ and\ \citenamefont
  {Jacobsen}}]{Sing03}%
  \BibitemOpen
  \bibfield  {author} {\bibinfo {author} {\bibfnamefont {M.}~\bibnamefont
  {Sing}}, \bibinfo {author} {\bibfnamefont {U.}~\bibnamefont
  {Schwingenschl\"ogl}}, \bibinfo {author} {\bibfnamefont {R.}~\bibnamefont
  {Claessen}}, \bibinfo {author} {\bibfnamefont {P.}~\bibnamefont {Blaha}},
  \bibinfo {author} {\bibfnamefont {J.~M.~P.}\ \bibnamefont {Carmelo}},
  \bibinfo {author} {\bibfnamefont {L.~M.}\ \bibnamefont {Martelo}}, \bibinfo
  {author} {\bibfnamefont {P.~D.}\ \bibnamefont {Sacramento}}, \bibinfo
  {author} {\bibfnamefont {M.}~\bibnamefont {Dressel}}, \ and\ \bibinfo
  {author} {\bibfnamefont {C.~S.}\ \bibnamefont {Jacobsen}},\ }\href {\doibase
  10.1103/PhysRevB.68.125111} {\bibfield  {journal} {\bibinfo  {journal} {Phys.
  Rev. B}\ }\textbf {\bibinfo {volume} {68}},\ \bibinfo {pages} {125111}
  (\bibinfo {year} {2003})}\BibitemShut {NoStop}%
\bibitem [{\citenamefont {Kim}\ \emph {et~al.}(2006)\citenamefont {Kim},
  \citenamefont {Koh}, \citenamefont {Rotenberg}, \citenamefont {Oh},
  \citenamefont {Eisaki}, \citenamefont {Motoyama}, \citenamefont {Uchida},
  \citenamefont {Tohyama}, \citenamefont {Maekawa}, \citenamefont {Shen},\ and\
  \citenamefont {Kim}}]{Kim2006}%
  \BibitemOpen
  \bibfield  {author} {\bibinfo {author} {\bibfnamefont {B.~J.}\ \bibnamefont
  {Kim}}, \bibinfo {author} {\bibfnamefont {H.}~\bibnamefont {Koh}}, \bibinfo
  {author} {\bibfnamefont {E.}~\bibnamefont {Rotenberg}}, \bibinfo {author}
  {\bibfnamefont {S.~J.}\ \bibnamefont {Oh}}, \bibinfo {author} {\bibfnamefont
  {H.}~\bibnamefont {Eisaki}}, \bibinfo {author} {\bibfnamefont
  {N.}~\bibnamefont {Motoyama}}, \bibinfo {author} {\bibfnamefont
  {S.}~\bibnamefont {Uchida}}, \bibinfo {author} {\bibfnamefont
  {T.}~\bibnamefont {Tohyama}}, \bibinfo {author} {\bibfnamefont
  {S.}~\bibnamefont {Maekawa}}, \bibinfo {author} {\bibfnamefont {Z.~X.}\
  \bibnamefont {Shen}}, \ and\ \bibinfo {author} {\bibfnamefont
  {C.}~\bibnamefont {Kim}},\ }\href {\doibase 10.1038/nphys316} {\bibfield
  {journal} {\bibinfo  {journal} {Nature Physics}\ }\textbf {\bibinfo {volume}
  {2}},\ \bibinfo {pages} {397} (\bibinfo {year} {2006})}\BibitemShut {NoStop}%
\bibitem [{\citenamefont {Sorella}\ and\ \citenamefont
  {Parola}(1992)}]{Sorella92}%
  \BibitemOpen
  \bibfield  {author} {\bibinfo {author} {\bibfnamefont {S.}~\bibnamefont
  {Sorella}}\ and\ \bibinfo {author} {\bibfnamefont {A.}~\bibnamefont
  {Parola}},\ }\href {\doibase 10.1088/0953-8984/4/13/020} {\bibfield
  {journal} {\bibinfo  {journal} {Journal of Physics: Condensed Matter}\
  }\textbf {\bibinfo {volume} {4}},\ \bibinfo {pages} {3589} (\bibinfo {year}
  {1992})}\BibitemShut {NoStop}%
\bibitem [{\citenamefont {Penc}\ \emph {et~al.}(1995)\citenamefont {Penc},
  \citenamefont {Mila},\ and\ \citenamefont {Shiba}}]{Penc95}%
  \BibitemOpen
  \bibfield  {author} {\bibinfo {author} {\bibfnamefont {K.}~\bibnamefont
  {Penc}}, \bibinfo {author} {\bibfnamefont {F.}~\bibnamefont {Mila}}, \ and\
  \bibinfo {author} {\bibfnamefont {H.}~\bibnamefont {Shiba}},\ }\href
  {\doibase 10.1103/PhysRevLett.75.894} {\bibfield  {journal} {\bibinfo
  {journal} {Phys. Rev. Lett.}\ }\textbf {\bibinfo {volume} {75}},\ \bibinfo
  {pages} {894} (\bibinfo {year} {1995})}\BibitemShut {NoStop}%
\bibitem [{\citenamefont {Penc}\ \emph {et~al.}(1996)\citenamefont {Penc},
  \citenamefont {Hallberg}, \citenamefont {Mila},\ and\ \citenamefont
  {Shiba}}]{Penc96}%
  \BibitemOpen
  \bibfield  {author} {\bibinfo {author} {\bibfnamefont {K.}~\bibnamefont
  {Penc}}, \bibinfo {author} {\bibfnamefont {K.}~\bibnamefont {Hallberg}},
  \bibinfo {author} {\bibfnamefont {F.}~\bibnamefont {Mila}}, \ and\ \bibinfo
  {author} {\bibfnamefont {H.}~\bibnamefont {Shiba}},\ }\href {\doibase
  10.1103/PhysRevLett.77.1390} {\bibfield  {journal} {\bibinfo  {journal}
  {Phys. Rev. Lett.}\ }\textbf {\bibinfo {volume} {77}},\ \bibinfo {pages}
  {1390} (\bibinfo {year} {1996})}\BibitemShut {NoStop}%
\bibitem [{\citenamefont {Penc}\ \emph {et~al.}(1997)\citenamefont {Penc},
  \citenamefont {Hallberg}, \citenamefont {Mila},\ and\ \citenamefont
  {Shiba}}]{Penc97}%
  \BibitemOpen
  \bibfield  {author} {\bibinfo {author} {\bibfnamefont {K.}~\bibnamefont
  {Penc}}, \bibinfo {author} {\bibfnamefont {K.}~\bibnamefont {Hallberg}},
  \bibinfo {author} {\bibfnamefont {F.}~\bibnamefont {Mila}}, \ and\ \bibinfo
  {author} {\bibfnamefont {H.}~\bibnamefont {Shiba}},\ }\href {\doibase
  10.1103/PhysRevB.55.15475} {\bibfield  {journal} {\bibinfo  {journal} {Phys.
  Rev. B}\ }\textbf {\bibinfo {volume} {55}},\ \bibinfo {pages} {15475}
  (\bibinfo {year} {1997})}\BibitemShut {NoStop}%
\bibitem [{\citenamefont {Tohyama}\ and\ \citenamefont
  {Maekawa}(1998)}]{TOHYAMA98}%
  \BibitemOpen
  \bibfield  {author} {\bibinfo {author} {\bibfnamefont {T.}~\bibnamefont
  {Tohyama}}\ and\ \bibinfo {author} {\bibfnamefont {S.}~\bibnamefont
  {Maekawa}},\ }\href {\doibase https://doi.org/10.1016/S0022-3697(98)00126-7}
  {\bibfield  {journal} {\bibinfo  {journal} {Journal of Physics and Chemistry
  of Solids}\ }\textbf {\bibinfo {volume} {59}},\ \bibinfo {pages} {1864}
  (\bibinfo {year} {1998})}\BibitemShut {NoStop}%
\bibitem [{\citenamefont {Aichhorn}\ \emph {et~al.}(2004)\citenamefont
  {Aichhorn}, \citenamefont {Evertz}, \citenamefont {von~der Linden},\ and\
  \citenamefont {Potthoff}}]{Aichhorn04}%
  \BibitemOpen
  \bibfield  {author} {\bibinfo {author} {\bibfnamefont {M.}~\bibnamefont
  {Aichhorn}}, \bibinfo {author} {\bibfnamefont {H.~G.}\ \bibnamefont
  {Evertz}}, \bibinfo {author} {\bibfnamefont {W.}~\bibnamefont {von~der
  Linden}}, \ and\ \bibinfo {author} {\bibfnamefont {M.}~\bibnamefont
  {Potthoff}},\ }\href {\doibase 10.1103/PhysRevB.70.235107} {\bibfield
  {journal} {\bibinfo  {journal} {Phys. Rev. B}\ }\textbf {\bibinfo {volume}
  {70}},\ \bibinfo {pages} {235107} (\bibinfo {year} {2004})}\BibitemShut
  {NoStop}%
\bibitem [{\citenamefont {Shao}\ \emph {et~al.}(2020)\citenamefont {Shao},
  \citenamefont {Tohyama}, \citenamefont {Luo},\ and\ \citenamefont
  {Lu}}]{Shao20}%
  \BibitemOpen
  \bibfield  {author} {\bibinfo {author} {\bibfnamefont {C.}~\bibnamefont
  {Shao}}, \bibinfo {author} {\bibfnamefont {T.}~\bibnamefont {Tohyama}},
  \bibinfo {author} {\bibfnamefont {H.-G.}\ \bibnamefont {Luo}}, \ and\
  \bibinfo {author} {\bibfnamefont {H.}~\bibnamefont {Lu}},\ }\href {\doibase
  10.1103/PhysRevB.101.045128} {\bibfield  {journal} {\bibinfo  {journal}
  {Phys. Rev. B}\ }\textbf {\bibinfo {volume} {101}},\ \bibinfo {pages}
  {045128} (\bibinfo {year} {2020})}\BibitemShut {NoStop}%
\bibitem [{\citenamefont {Tsuchiizu}\ and\ \citenamefont
  {Furusaki}(2004)}]{Tsuchiizu04}%
  \BibitemOpen
  \bibfield  {author} {\bibinfo {author} {\bibfnamefont {M.}~\bibnamefont
  {Tsuchiizu}}\ and\ \bibinfo {author} {\bibfnamefont {A.}~\bibnamefont
  {Furusaki}},\ }\href {\doibase 10.1103/PhysRevB.69.035103} {\bibfield
  {journal} {\bibinfo  {journal} {Phys. Rev. B}\ }\textbf {\bibinfo {volume}
  {69}},\ \bibinfo {pages} {035103} (\bibinfo {year} {2004})}\BibitemShut
  {NoStop}%
\bibitem [{\citenamefont {Ejima}\ \emph {et~al.}(2016)\citenamefont {Ejima},
  \citenamefont {Essler}, \citenamefont {Lange},\ and\ \citenamefont
  {Fehske}}]{Ejima16}%
  \BibitemOpen
  \bibfield  {author} {\bibinfo {author} {\bibfnamefont {S.}~\bibnamefont
  {Ejima}}, \bibinfo {author} {\bibfnamefont {F.~H.~L.}\ \bibnamefont
  {Essler}}, \bibinfo {author} {\bibfnamefont {F.}~\bibnamefont {Lange}}, \
  and\ \bibinfo {author} {\bibfnamefont {H.}~\bibnamefont {Fehske}},\ }\href
  {\doibase 10.1103/PhysRevB.93.235118} {\bibfield  {journal} {\bibinfo
  {journal} {Phys. Rev. B}\ }\textbf {\bibinfo {volume} {93}},\ \bibinfo
  {pages} {235118} (\bibinfo {year} {2016})}\BibitemShut {NoStop}%
\bibitem [{\citenamefont {Le}\ \emph {et~al.}(2020)\citenamefont {Le},
  \citenamefont {Fisher}, \citenamefont {Curson},\ and\ \citenamefont
  {Ginossar}}]{Le2020}%
  \BibitemOpen
  \bibfield  {author} {\bibinfo {author} {\bibfnamefont {N.~H.}\ \bibnamefont
  {Le}}, \bibinfo {author} {\bibfnamefont {A.~J.}\ \bibnamefont {Fisher}},
  \bibinfo {author} {\bibfnamefont {N.~J.}\ \bibnamefont {Curson}}, \ and\
  \bibinfo {author} {\bibfnamefont {E.}~\bibnamefont {Ginossar}},\ }\href
  {\doibase 10.1038/s41534-020-0253-9} {\bibfield  {journal} {\bibinfo
  {journal} {npj Quantum Information}\ }\textbf {\bibinfo {volume} {6}},\
  \bibinfo {pages} {24} (\bibinfo {year} {2020})}\BibitemShut {NoStop}%
\bibitem [{\citenamefont {Pedron}\ \emph {et~al.}(1994)\citenamefont {Pedron},
  \citenamefont {Bozio}, \citenamefont {Meneghetti},\ and\ \citenamefont
  {Pecile}}]{Pedron94}%
  \BibitemOpen
  \bibfield  {author} {\bibinfo {author} {\bibfnamefont {D.}~\bibnamefont
  {Pedron}}, \bibinfo {author} {\bibfnamefont {R.}~\bibnamefont {Bozio}},
  \bibinfo {author} {\bibfnamefont {M.}~\bibnamefont {Meneghetti}}, \ and\
  \bibinfo {author} {\bibfnamefont {C.}~\bibnamefont {Pecile}},\ }\href
  {\doibase 10.1103/PhysRevB.49.10893} {\bibfield  {journal} {\bibinfo
  {journal} {Phys. Rev. B}\ }\textbf {\bibinfo {volume} {49}},\ \bibinfo
  {pages} {10893} (\bibinfo {year} {1994})}\BibitemShut {NoStop}%
\bibitem [{\citenamefont {Nishimoto}\ \emph {et~al.}(2000)\citenamefont
  {Nishimoto}, \citenamefont {Takahashi},\ and\ \citenamefont
  {Ohta}}]{Nishimoto2000}%
  \BibitemOpen
  \bibfield  {author} {\bibinfo {author} {\bibfnamefont {S.}~\bibnamefont
  {Nishimoto}}, \bibinfo {author} {\bibfnamefont {M.}~\bibnamefont
  {Takahashi}}, \ and\ \bibinfo {author} {\bibfnamefont {Y.}~\bibnamefont
  {Ohta}},\ }\href {\doibase 10.1143/jpsj.69.1594} {\bibfield  {journal}
  {\bibinfo  {journal} {Journal of the Physical Society of Japan}\ }\textbf
  {\bibinfo {volume} {69}},\ \bibinfo {pages} {1594} (\bibinfo {year}
  {2000})}\BibitemShut {NoStop}%
\bibitem [{\citenamefont {Shibata}\ \emph {et~al.}(2001)\citenamefont
  {Shibata}, \citenamefont {Nishimoto},\ and\ \citenamefont
  {Ohta}}]{Shibata01}%
  \BibitemOpen
  \bibfield  {author} {\bibinfo {author} {\bibfnamefont {Y.}~\bibnamefont
  {Shibata}}, \bibinfo {author} {\bibfnamefont {S.}~\bibnamefont {Nishimoto}},
  \ and\ \bibinfo {author} {\bibfnamefont {Y.}~\bibnamefont {Ohta}},\ }\href
  {\doibase 10.1103/PhysRevB.64.235107} {\bibfield  {journal} {\bibinfo
  {journal} {Phys. Rev. B}\ }\textbf {\bibinfo {volume} {64}},\ \bibinfo
  {pages} {235107} (\bibinfo {year} {2001})}\BibitemShut {NoStop}%
\bibitem [{\citenamefont {Tsuchiizu}\ \emph {et~al.}(2001)\citenamefont
  {Tsuchiizu}, \citenamefont {Yoshioka},\ and\ \citenamefont
  {Suzumura}}]{Tsuchiizu01}%
  \BibitemOpen
  \bibfield  {author} {\bibinfo {author} {\bibfnamefont {M.}~\bibnamefont
  {Tsuchiizu}}, \bibinfo {author} {\bibfnamefont {H.}~\bibnamefont {Yoshioka}},
  \ and\ \bibinfo {author} {\bibfnamefont {Y.}~\bibnamefont {Suzumura}},\
  }\href {\doibase 10.1143/JPSJ.70.1460} {\bibfield  {journal} {\bibinfo
  {journal} {Journal of the Physical Society of Japan}\ }\textbf {\bibinfo
  {volume} {70}},\ \bibinfo {pages} {1460} (\bibinfo {year}
  {2001})}\BibitemShut {NoStop}%
\bibitem [{\citenamefont {Penc}\ and\ \citenamefont {Mila}(1994)}]{Penc94}%
  \BibitemOpen
  \bibfield  {author} {\bibinfo {author} {\bibfnamefont {K.}~\bibnamefont
  {Penc}}\ and\ \bibinfo {author} {\bibfnamefont {F.}~\bibnamefont {Mila}},\
  }\href {\doibase 10.1103/PhysRevB.50.11429} {\bibfield  {journal} {\bibinfo
  {journal} {Phys. Rev. B}\ }\textbf {\bibinfo {volume} {50}},\ \bibinfo
  {pages} {11429} (\bibinfo {year} {1994})}\BibitemShut {NoStop}%
\bibitem [{\citenamefont {Mila}(1995)}]{Mila95}%
  \BibitemOpen
  \bibfield  {author} {\bibinfo {author} {\bibfnamefont {F.}~\bibnamefont
  {Mila}},\ }\href {\doibase 10.1103/PhysRevB.52.4788} {\bibfield  {journal}
  {\bibinfo  {journal} {Phys. Rev. B}\ }\textbf {\bibinfo {volume} {52}},\
  \bibinfo {pages} {4788} (\bibinfo {year} {1995})}\BibitemShut {NoStop}%
\bibitem [{\citenamefont {Favand}\ and\ \citenamefont {Mila}(1996)}]{Favand96}%
  \BibitemOpen
  \bibfield  {author} {\bibinfo {author} {\bibfnamefont {J.}~\bibnamefont
  {Favand}}\ and\ \bibinfo {author} {\bibfnamefont {F.}~\bibnamefont {Mila}},\
  }\href {\doibase 10.1103/PhysRevB.54.10425} {\bibfield  {journal} {\bibinfo
  {journal} {Phys. Rev. B}\ }\textbf {\bibinfo {volume} {54}},\ \bibinfo
  {pages} {10425} (\bibinfo {year} {1996})}\BibitemShut {NoStop}%
\bibitem [{\citenamefont {Benthien}\ and\ \citenamefont
  {Jeckelmann}(2005)}]{Benthien2005}%
  \BibitemOpen
  \bibfield  {author} {\bibinfo {author} {\bibfnamefont {H.}~\bibnamefont
  {Benthien}}\ and\ \bibinfo {author} {\bibfnamefont {E.}~\bibnamefont
  {Jeckelmann}},\ }\href {\doibase 10.1140/epjb/e2005-00128-1} {\bibfield
  {journal} {\bibinfo  {journal} {The European Physical Journal B - Condensed
  Matter and Complex Systems}\ }\textbf {\bibinfo {volume} {44}},\ \bibinfo
  {pages} {287} (\bibinfo {year} {2005})}\BibitemShut {NoStop}%
\bibitem [{\citenamefont {Tsutsui}\ \emph {et~al.}(1996)\citenamefont
  {Tsutsui}, \citenamefont {Ohta}, \citenamefont {Eder}, \citenamefont
  {Maekawa}, \citenamefont {Dagotto},\ and\ \citenamefont {Riera}}]{Tsutsui96}%
  \BibitemOpen
  \bibfield  {author} {\bibinfo {author} {\bibfnamefont {K.}~\bibnamefont
  {Tsutsui}}, \bibinfo {author} {\bibfnamefont {Y.}~\bibnamefont {Ohta}},
  \bibinfo {author} {\bibfnamefont {R.}~\bibnamefont {Eder}}, \bibinfo {author}
  {\bibfnamefont {S.}~\bibnamefont {Maekawa}}, \bibinfo {author} {\bibfnamefont
  {E.}~\bibnamefont {Dagotto}}, \ and\ \bibinfo {author} {\bibfnamefont
  {J.}~\bibnamefont {Riera}},\ }\href {\doibase 10.1103/PhysRevLett.76.279}
  {\bibfield  {journal} {\bibinfo  {journal} {Phys. Rev. Lett.}\ }\textbf
  {\bibinfo {volume} {76}},\ \bibinfo {pages} {279} (\bibinfo {year}
  {1996})}\BibitemShut {NoStop}%
\bibitem [{\citenamefont {Tohyama}(2004)}]{Tohyama04}%
  \BibitemOpen
  \bibfield  {author} {\bibinfo {author} {\bibfnamefont {T.}~\bibnamefont
  {Tohyama}},\ }\href {\doibase 10.1103/PhysRevB.70.174517} {\bibfield
  {journal} {\bibinfo  {journal} {Phys. Rev. B}\ }\textbf {\bibinfo {volume}
  {70}},\ \bibinfo {pages} {174517} (\bibinfo {year} {2004})}\BibitemShut
  {NoStop}%
\bibitem [{\citenamefont {Su}\ \emph {et~al.}(2023)\citenamefont {Su},
  \citenamefont {Lu}, \citenamefont {Lu},\ and\ \citenamefont
  {Shao}}]{Su_2023}%
  \BibitemOpen
  \bibfield  {author} {\bibinfo {author} {\bibfnamefont {Y.-G.}\ \bibnamefont
  {Su}}, \bibinfo {author} {\bibfnamefont {R.}~\bibnamefont {Lu}}, \bibinfo
  {author} {\bibfnamefont {H.}~\bibnamefont {Lu}}, \ and\ \bibinfo {author}
  {\bibfnamefont {C.}~\bibnamefont {Shao}},\ }\href {\doibase
  10.1088/1361-6455/acc49b} {\bibfield  {journal} {\bibinfo  {journal} {Journal
  of Physics B: Atomic, Molecular and Optical Physics}\ }\textbf {\bibinfo
  {volume} {56}},\ \bibinfo {pages} {085101} (\bibinfo {year}
  {2023})}\BibitemShut {NoStop}%
\end{thebibliography}

%

\end{document}